\documentclass[journal=jacsat,manuscript=article]{achemso}

\setkeys{acs}{articletitle=true,etalmode=truncate,maxauthors=5}

\usepackage{caption}
\usepackage{float}
\usepackage{geometry}
\usepackage{xkeyval}
\usepackage{setspace}
\usepackage{tabularx}
\usepackage{graphicx}
\usepackage{dcolumn}
\usepackage{bm}
\usepackage{soul}
\usepackage{color}
\usepackage{xcolor}
\usepackage{amsmath}
\usepackage{amssymb}
\usepackage{amsmath}
\usepackage{natmove}
\usepackage{multirow}
\usepackage{array}
\usepackage{booktabs}
\usepackage{rotating}
\usepackage[normalem]{ulem}
\usepackage[colorlinks=true,linkcolor=blue,citecolor=blue,urlcolor=blue]{hyperref}
\usepackage{physics}
\usepackage{mathrsfs}
\usepackage[capitalise,noabbrev]{cleveref}
\usepackage[version=3]{mhchem} 
\usepackage{comment}
\usepackage{pdfpages}

\definecolor{Grey}{rgb}{0.50,0.50,0.50}
\definecolor{Blue}{rgb}{0.00,0.00,1.00}
\definecolor{Red}{rgb}{1.00,0.00,0.00}
\definecolor{Green}{rgb}{0.20,0.80,0.20}
\definecolor{Magenta}{rgb}{0.60,0.00,0.60}
\definecolor{BluBondi}{rgb}{0.00,0.58,0.71}
\definecolor{Orange}{rgb}{0.95,0.46,0.17}

\newcommand{\editor}[2]{%
  \expandafter\newcommand\csname #1note\endcsname[1]{%
    \textcolor{#2}{{\it (\textbf{#1:} ##1)}}}%
  \expandafter\newcommand\csname #1\endcsname[1]{%
    \textcolor{#2}{##1}}%
  \expandafter\newcommand\csname #1cancel\endcsname[1]{%
    \textcolor{#2}{\sout{##1}}}%
  \expandafter\newcommand\csname #1change\endcsname[2]{%
    \textcolor{#2}{\sout{##1} ##2}}%
  \newenvironment{#1text}{\color{#2}}{\color{black}}
}

\editor{DP}{cyan}
\editor{MH}{Blue}
\editor{DZ}{orange}
\editor{SL}{red}
\editor{AL}{Green}
\editor{EM}{magenta}

\newcommand{\drop}[1]{}

\author{Michael A. Hernandez Bertran}
\affiliation{Physics, Informatics and Mathematics Department (FIM), University of Modena and Reggio Emilia, I-41125 Modena, Italy}
\alsoaffiliation{Nanoscience Institute -- National Research Council (CNR-NANO), I-41125 Modena, Italy}
\altaffiliation{These authors contributed equally to this work}

\author{Diana Zapata Dominguez}
\affiliation{University Grenoble Alpes, CEA, IRIG, MEM, F-38054 Grenoble, France}
\altaffiliation{These authors contributed equally to this work}

\author{Christopher Berhaut}
\affiliation{University Grenoble Alpes, CEA, CNRS, IRIG-SyMMES, F-38054 Grenoble, France}

\author{Samuel Tardif}
\affiliation{University Grenoble Alpes, CEA, IRIG, MEM, F-38054 Grenoble, France}

\author{Alessandro Longo}
\affiliation{ESRF, The European Synchrotron, CS40220, Cedex 9, F-38043 Grenoble, France}

\author{Christoph Sahle}
\affiliation{ESRF, The European Synchrotron, CS40220, Cedex 9, F-38043 Grenoble, France}

\author{Chiara Cavallari}
\affiliation{ESRF, The European Synchrotron, CS40220, Cedex 9, F-38043 Grenoble, France}

\author{Ivan Marri}
\affiliation{Department of Sciences and Methods for Engineering, University of Modena and Reggio Emilia, I-42122 Reggio Emilia, Italy}

\author{Nathalie Herlin-Boime}
\affiliation{Universit\'{e} Paris-Saclay, CNRS, CEA-Saclay, NIMBE, UMR 3685 CEA, F-91191, Gif-sur-Yvette Cedex, France}

\author{Elisa Molinari}
\affiliation{Physics, Informatics and Mathematics Department (FIM), University of Modena and Reggio Emilia, I-41125 Modena, Italy}
\alsoaffiliation{Nanoscience Institute -- National Research Council (CNR-NANO), I-41125 Modena, Italy}

\author{St\'{e}phanie Pouget}
\affiliation{University Grenoble Alpes, CEA, IRIG, MEM, F-38054 Grenoble, France}

\author{Deborah Prezzi}
\affiliation{Nanoscience Institute -- National Research Council (CNR-NANO), I-41125 Modena, Italy}
\email{deborah.prezzi@nano.cnr.it}

\author{Sandrine Lyonnard}
\affiliation{University Grenoble Alpes, CEA, CNRS, IRIG, SyMMES, F-38054 Grenoble, France}
\email{sandrine.lyonnard@cea.fr}

\title{Understanding the irreversible lithium loss in silicon anodes using multi-edge X-ray scattering analysis}

\abbreviations{}
\keywords{X-ray Raman spectroscopy, density functional theory, silicon electrodes, solid electrolyte interphase}

\begin{document}

%

\newpage

\begin{abstract}
During the first charge-discharge cycle, silicon-based batteries show an important capacity loss because of the formation of the solid electrolyte interphase (SEI) and morphological changes due to expansion-contraction sequence upon alloying. To understand this first-cycle irreversibility, quantitative methods are needed to characterize the chemical environment of silicon and lithium in the bulk of the cycled electrodes. Here we report a methodology based on multi-edge X-ray Raman Scattering performed on model silicon electrodes prepared in fully lithiated and fully delithiated states after the first cycle. The spectra were recorded at the C, O, F and Li K edges, as well as Si \ce{L_2,3} edge. They were analysed using linear combinations of both experimental and computed reference spectra. We used prototypical SEI compounds as \ce{Li_2CO_3}, \ce{LiF} and \ce{LiPF_6}, as well as electrode constituents as binder and conductive carbon, cristalline Si, native \ce{SiO2}, \ce{Li_xSi} phases (x being the lithiation index) to identify the main species, isolate their relative contributions, and quantitatively evaluate the proportions of organic and inorganic products. We find that 30\% of the carbonates formed in the SEI during the lithiation are dissolved on delithiation, and that part of the \ce{Li_15Si_4} alloys remain present after delithiation. By combining electrochemical analysis and XRS results, we identify that 17\% of the lithium lost in the first cycle is trapped in disconnected silicon particles, while 30\% form a fluorine-rich stable SEI and 53\% a carbonate-rich partially-dissolvable SEI. These results pave the way to systematic, reference data-informed, and modelling assisted studies of SEI characteristics in the bulk of electrodes prepared under controlled state-of-charge and state-of-health conditions. 
\end{abstract}

\section{Introduction}

Lithium-ion batteries (LIB) degrade over time due to various physical and chemical phenomena, leading to irreversible capacity loss. Mitigating this aging process is crucial to extend the lifespan of electrochemical cells, meeting societal demands for a more sustainable energy production and storage chain. These requirements add on top of the quest for high-energy-density and cost-effective batteries, which is key for the success of electric vehicles and large grid storage.

Achieving these ambitious goals in terms of performance and sustainability requires significant advancements in materials chemistry, component manufacturing, and cell fabrication, especially for state-of-the-art technologies like silicon-graphite/NMC (or LFP) cells~\cite{yourey2023,gasper2023}, which currently dominate the market. In these cells, silicon is integrated in graphite-based composites to improve performances, given its ten times higher theoretical gravimetric capacity \cite{wu2012,eshetu2021,yang2023}. However, the amount of Si has generally to be limited to few weight percent (wt\%) to avoid quick performance decay due to loss of active material, loss of cyclable lithium and other degradation mechanisms \cite{chae2020}. 
Indeed, silicon suffers from severe capacity fading issues connected both to the drastic volume expansion experienced when alloying with Li \cite{lu2015,michan2016,he2021}, and to the formation of a dynamic solid electrolyte interphase (SEI) \cite{michan2016,he2021}
, which continuously evolves during swelling/shrinking of the silicon material, further contributing to additional particle cracking, pulverization and disconnections~\cite{taiwo2017,muller2020,muller2018} as well as Li loss~\cite{kumar2020}.


Mitigating these effects can be achieved by using amorphous, nanoporous and/or nanostructured Si \cite{liu2012y,kim2014,choi2017,liu2014,an2019}, together with the usage of polymeric binders~\cite{hochgatterer2008,vogl2014} and the design of a stable SEI, which is usually obtained by the use of additives as fluoroethylene carbonate (FEC) and vinylene carbonate (VC)~\cite{nie2013,martinez2014,ma2014,michan2016,jin2018,li2019}.
Other directions to prevent lithium loss is to use prelithiation \cite{wang2021prelithiation} 
or to blend silicon with another group IV material \cite{zapata2023}.
Nevertheless, effective control-by-design strategies are limited by the difficulty to understand the origin of lithium loss in silicon compounds. This is partly due to the difficulty to probe the SEI characteristics as a function of electrode/electrolyte composition and cycling parameters, as well as to localize and quantify lithium trapped in the cycled structures at particle, electrode and cell scales \cite{petz2021}.

Regarding the SEI, its interfacial nature, complex composition, variable thickness and evolution over cycling, represent a true challenge for experimental characterization. Techniques such as nuclear magnetic resonance (NMR)~\cite{michan2016,hu2019,michan2016cm2}, X-ray photoelectron spectroscopy (XPS)~\cite{xu2015,xu2014,philippe2012,philippe2013,hernandez2020}, Raman spectroscopy~\cite{nanda2019}, Fourier transform infrared spectroscopy (FTIR)~\cite{ruther2018}, and scanning transmission electron microscopy with energy-loss spectroscopy (STEM-EELS)~\cite{boniface2016} have provided insights into the SEI composition, primarily through post-mortem analysis. It is now widely recognized that the SEI is mainly composed of \ce{Li2CO3}, \ce{LiF}, oligomer species and semicarbonates~\cite{gauthier2015}, in addition to other (in)organic compounds specific to the electrode and electrolyte composition, as well as to the usage of additives~\cite{xu2015,li2019,michan2016,jin2018}. However, fundamental questions remain regarding the organic vs inorganic compounds distribution, as well as their relative role and dynamic evolution during cycling. For instance, some authors found that LiF is quite stable over time, forming thin film patches that cover the surface of silicon particles \cite{kumar2020}, 
while others report that LiF is dissolved partially during cycling \cite{desrues2022}. 
Many of these studies are limited to surface analysis (typically, probing few nanometres) and can be affected by surface contamination or intrusive sample preparation, when performed \emph{ex situ}. Alternatively, operando characterizations are attractive because they enable the comparison between different states-of-charge of the same sample without dismounting the cells and manipulating the electrodes, using for instance standard spectroscopies as XAS or NMR, or reflectivity techniques \cite{cao2019}, 
but they are not often conclusive or representative due to specific set-ups, beam damage issues, or lack of quantification\cite{swallow2022,schellenberger2022}. 

Beyond SEI studies, accessing lithium inventory in a cycling battery is also difficult. Combined synchrotron experiments coupling X-ray diffraction and X-ray imaging \cite{pietsch2016}, 
or X-ray diffraction and X-ray fluorescence \cite{dawkins2023}, 
have attempted to identify and quantify the presence and distribution of \ce{Li_xSi} phases (x being the lithiation index) from analysing spectral signatures, combining it with total lithium content analysis to deconvolute the various degradation processes leading to lithium loss. However, these studies remain scarce and specific, calling for the need of dedicated bulk-level chemical-sensitive investigation in model systems to establish the nature and proportions of ions lost in silicon, particularly during the first cycle where the SEI forms. 

Here we fill this gap by applying synchrotron X-ray Raman Scattering (XRS) measurements and multi-edge analysis via a combined experimental-numerical approach. Indeed, XRS offers a promising alternative to evaluate both the mechanism of lithium loss in silicon anodes and the SEI characteristics, at it provides information on the bulk electronic structures and chemical environments of materials~\cite{ketenoglu2018,fehse2021}. Unlike absorption techniques, XRS is a non-resonant technique that uses highly penetrating hard X-rays, which are inelastically scattered by the material allowing access to soft X-ray edges in the bulk. This enables detailed analysis of light elements like Li, C, O, and F, which are crucial components of the SEI, as well as access to Si edges (both K and L$_{2,3}$), hence providing information on silicon alloys and oxides that are formed and evolve within the electrode upon cycling. 

To date, XRS has had limited application in battery research~\cite{braun2015,liu2012,nonaka2019,boesenberg2019,deboer2020, fehse2021,rajh2022,dekort2024,togonon2024} and, 
despite showing great potential, this technique is still in its infancy. One of the main issues is the quantitative analysis of XRS data, which remains challenging due to limited reference spectra as compared to more established XAS or XPS~\cite{fehse2021}. Moreover, in the absence of other references, the analysis of multi-component spectra requires numerical, DFT-aided computations to disentangle contributions in complex environments. Finally, the long measurement times required to optimize the signal-to-noise ratio make operando studies limited to restricted configurations and conditions at present~\cite{ketenoglu2018}, with real operando cells experiments remaining beyond reach for the moment. Therefore, detailed analysis of redox or degradation reactions in controlled and representative conditions by XRS require substantial improvements at the level of experimental capabilities, database construction, and modelling.    

In this work, we tackle the first two challenges by developing a multi-edge combined experimental-computational approach to analyse \emph{ex situ} X-ray Raman Scattering of silicon-based anodes. Series of high-resolution synchrotron XRS data are collected at the Li, C, O, F and Si edges in a benchmark silicon-nanoparticles pristine anode, as well as in its lithiated and delithiated states during the first cycle. XRS spectra of several key reference compounds are acquired and simulated to serve as benchmarks for data analysis. We develop the methodology to investigate the multi-component SEI layer composition and quantify lithium loss by using computed model spectra to support the linear deconvolution of the XRS data. 
With this approach, we identify the main SEI species, isolate their relative contributions, and quantitatively evaluate organic and inorganic products, such as \ce{Li2CO3} and \ce{LiF}. Additionally, we analyse the Si environment and detect the presence of residual \ce{Li_xSi} phases after delithiation, as confirmed by Li edge data. 


\section{Results and discussion}

We use a prototypical anode, i.e., a silicon electrode composed of crystalline nanoparticles (c-Si NPs), synthesized by laser pyrolysis \cite{desrues2019} with diameters between 60 and 100 nm, mixed with a binder (sodium carboxymethyl cellulose, Na-CMC) and a conductive agent (carbon black, hereafter labelled Super P or SP). 
The electrodes were cycled in half-cell using a standard carbonate-based electrolyte with \ce{LiPF6} salt and vinyl carbonate (VC) additive. 
Samples were prepared in two states of charge (SoC) during the first charge-discharge cycle performed at C/20, that is, fully lithiated (SoC = 100\%) and fully delithiated (SoC = 0\%), and the capacity loss of the first charge-discharge cycle was measured by means of standard galvanostatic measurements (see next Subsection). Further details on the materials and cycling protocols are reported in Materials and Methods section.

Without any prior rinsing, the two cycled electrodes were next measured by XRS at the ESRF synchrotron using a custom air-tight cell, together with the pristine electrode and pristine powder materials. A full energy scan was performed for each sample from 0 to 800 eV, followed by high resolution scans taken at all relevant edges, i.e., Li, C, O, F K-edge and Si L$_{2,3}$-edge. Details on set-up, measurement protocols, XRS data reduction and corrections, are given in the Methods, while all the acquired XRS data are represented on Figure S1 of the Supporting Information. At first glance, it is clear that the SoC affects the overall shapes and intensities of the electrodes spectra, some edges being more altered than others. Moreover, specific features can be visually recognized by comparing the silicon electrode data to the model materials, e.g., for instance specific low-energy C-fingerprint of \ce{Li2CO3}, highlighting the interest of the database-informed strategy. 

To quantify the observed changes and correlations, XRS results are analysed in the following by comparing the experimental spectra of the cycled electrodes obtained at the various edges to a linear fitting of reference spectra (see Supporting Information for the detailed fitting procedure) acquired on purpose in the same conditions, on the same set-up, using selected compounds of interest. Modelled XRS spectra (see Method section) are also used to support the attribution of the observed changes to specific species and products, as will be discussed in details in the next sections. Specifically, we next consider: (i) the C, O and F edges, which contain information on the main organic and inorganic SEI products; (ii) the Si edge data, in order to compare the presence of silicon-based alloys in the two SoC; (iii) the Li edge, allowing us to discuss the complex chemical environments present in the electrodes and the implications of the observed energy shifts and profiles variations in terms of lithium inventory. This multi-edge analysis enables us to integrate all numerical and experimental observations into a model of the lithium loss analysis, quantifying the different populations of lithium chemical environment. 

\subsection{Galvanostatic cycling measurement}
\label{subsec:galvanostatic}

\begin{figure}
    \centering
    \includegraphics[width=0.50\textwidth]{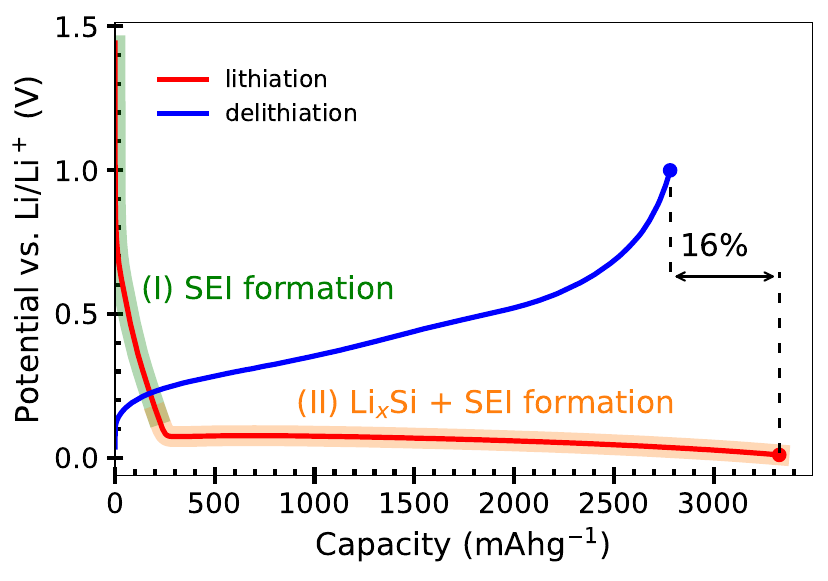}
\caption{Potential vs. Li/Li$^+$ against capacity during the first electrochemical cycle in a coin cell at C/20 for crystalline Si-NPs based anodes. The electrode SOC analyzed post-mortem by XRS are indicated by red/blue dots for the fully lithiated/delithiated state, respectively.}
\label{fig:galvanostatic}
\end{figure}

\cref{fig:galvanostatic} shows the first electrochemical cycle during lithiation up to 0.01 V versus Li metal in a coin cell. The electrochemical curve features a plateau at 0.08 V, similar to that previously reported for c-Si NPs~\cite{obrovac2004,obrovac2007,pinilla2020}. The discharge capacity obtained is 3328 mAhg$^{-1}$ (SoC = 100\%, red dot).
During delithiation, the obtained capacity reaches 2781 mAhg$^{-1}$ (SoC = 0\%, blue dot), with a Coulombic efficiency of~84\% and an irreversible capacity loss of~16\%, in agreement with previous results on c-Si NPs~\cite{bridel2010,philippe2012}. 

The capacity loss measured during the first cycle is different from any other, and usually related to the initial SEI formation observed for Si-based electrodes~\cite{philippe2012,radvanyi2014,kumar2020}. 
Indeed, during the first lithiation, two regions can be identified on the galvanostatic curve (marked in green and orange, respectively).
For potentials above $\sim 150$ mV (green), the irreversible reduction of solvents and salts in the electrolyte is known to take place, resulting in the formation of an initial, highly porous SEI, primarily composed of fluorinated compounds~\cite{chan2009,liu2014acs,desrues2022}. Below this lithiation potential (orange), Si lithiation starts to occur~\cite{obrovac2007,ogata2014,michan2016cm1} while the SEI evolution continues, with a major contribution from \ce{Li2CO3} \cite{chan2009,liu2014acs}. 
Following delithiation, part of the SEI decomposes and some amount of Li remains trapped inside the Si-NPs \cite{liu2005,zhao2020,sreenarayanan2022}.

\subsection{C, O and F K-edge: tracking the early SEI formation}\label{subsec:sei}

\begin{figure*}[h!]
    \centering
    \includegraphics[width=1.0\textwidth]{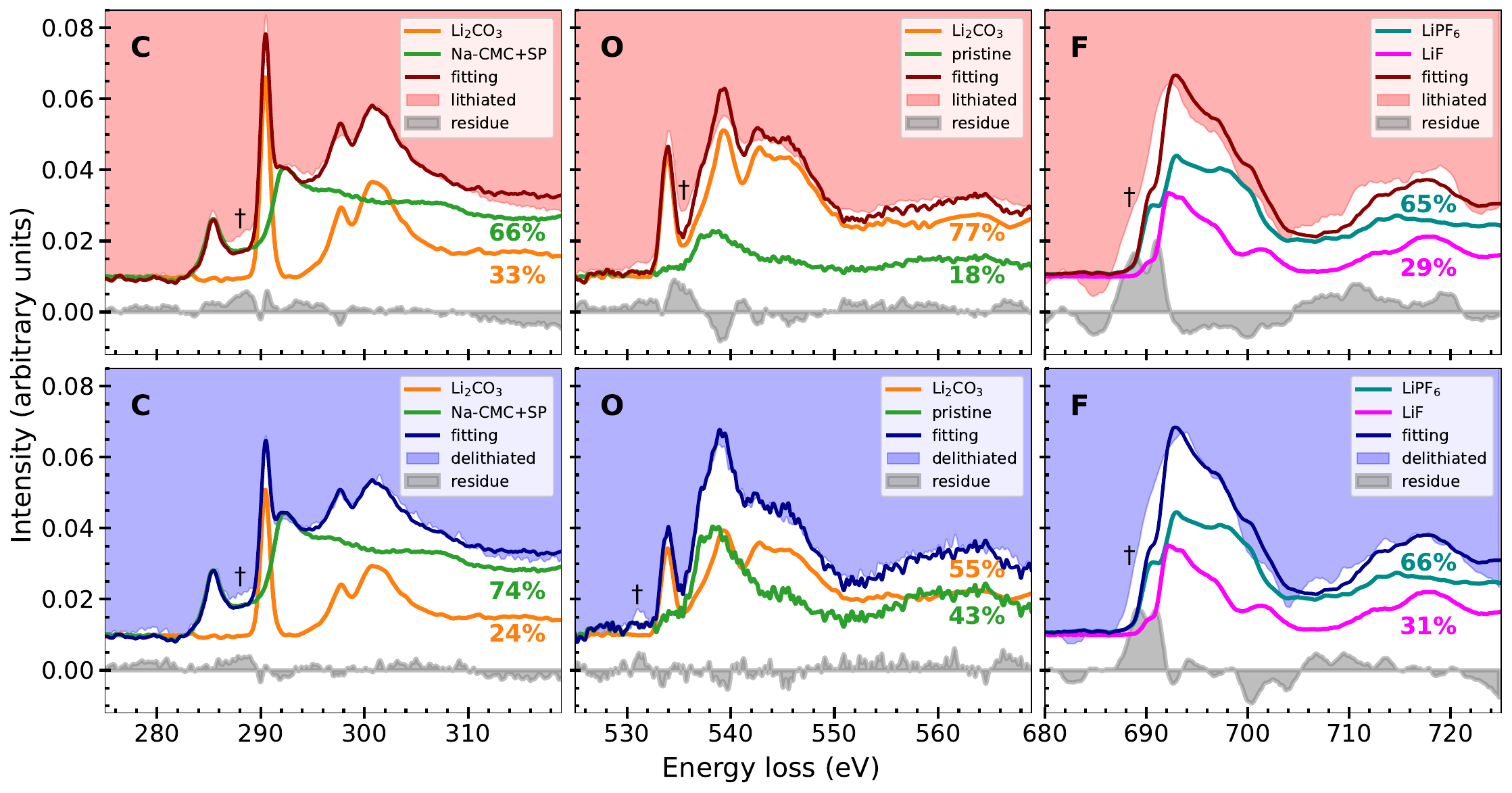}
\caption{Experimental non-resonant inelastic X-ray scattering spectra of c-Si NPs electrodes in the low momentum transfer regime compared with the linear combination of selected references for the C K-edge (left), O K-edge (center) and F K-edge (right) in the lithiated (top) and delithiated (bottom) states.
}\label{fig:XRSall}
\end{figure*}

\autoref{fig:XRSall} (white region) shows the C, O and F K-edge XRS spectra (normalized to unit area) of electrode samples at the end of the first charge (lithiation, top) and the first discharge (delithiation, bottom) sequence, which correspond to the two different SoC identified by red and blue dot in \cref{fig:galvanostatic}, respectively. The best fit obtained from the linear combination (LC) of relevant experimental reference spectra is also reported in \autoref{fig:XRSall} for each edge and each state of charge (dark red/blue curve for the lithiated/delithiated state), together with the weighted experimental references (other coloured solid curves with indicated \%) and the residual curve after fitting (grey area).
%
The reference compounds were selected according to the main decomposition products of the electrolyte in LIB (i.e., \ce{Li2CO3}, \ce{LiPF6} and \ce{LiF}) and the initial constituents of the pristine electrode (Na-CMC + Super P and Si-NPs). Note that, as the electrodes are not rinsed, \ce{LiPF6} is present in the liquid electrolyte filling the pores. 

Overall, the spectra of the Si NPs electrodes at different SOC are fairly well described by the LC model. In particular, only minor differences, accounted by the residual curve, are detected for the C and O (delithiated) K-edges. The energy regions where the fitted LC model deviates appreciably from the measurements ($\dagger$) reflect the incompleteness of the basis expansion. Deviations are more prominently observed in the O (lithiated) and F K-edges, and will be addressed in the following by resorting to density-functional theory (DFT) based computational modelling.

\begin{figure}[h!]
    \centering
    \includegraphics[width=0.5\textwidth]{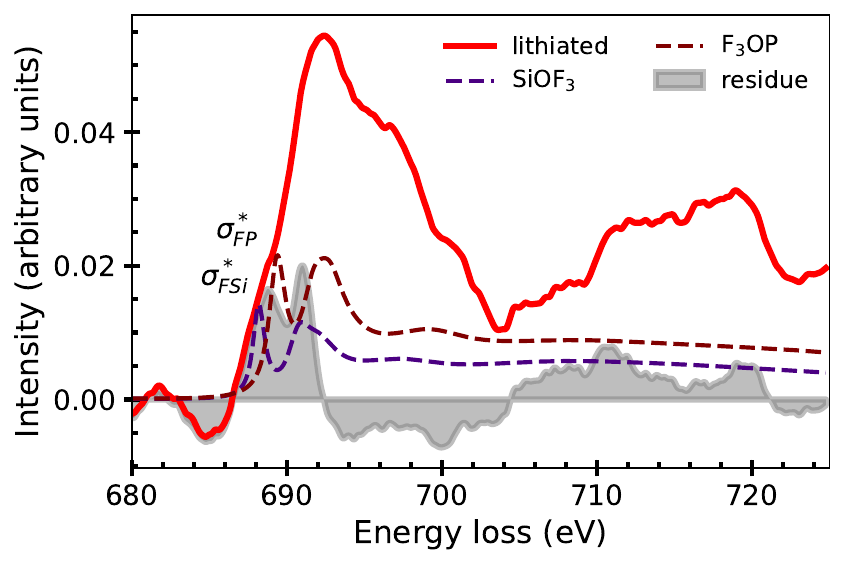}
\caption{XRS F K-edge spectrum of c-Si NPs electrodes in the low momentum transfer regime. The experimental curve in the lithiated state (red line) is compared to selected theoretical references (dashed lines). The residual curve from ~\cref{fig:XRSall} is also reported in grey.}\label{fig:XAS_F}
\end{figure}

\emph{F K-edge.} We start by analyzing the F edge (\cref{fig:XRSall}, right panels), reminding that the pristine electrode is F-free. We notice that, for both SoC, the best fit to the XRS spectrum requires the presence of both references, i.e. \ce{LiPF6} and \ce{LiF}, confirming 
the formation of \ce{LiF} already during the first cycle~\cite{michan2016}. 
Whereas the relative composition obtained by the fitting remains nearly unchanged (see \%) for the two SoC, we notice an upward shift in energy of 1.5 eV for the edge maximum located at 692.4 eV when the cell is delithiated. 
Moreover, 
%
we find for both SoC a lower energy onset at about 689 eV ($\dagger$) as compared to the selected references, suggesting the presence of additional F-bearing compounds. 

A previous work detected \ce{PO_xF_y} and \ce{SiO_xF_y} on similar systems using \textit{ex-situ} multi-nuclear solid-state NMR \ce{^19F} spectra and Cross Polarization (CP) experiments~\cite{michan2016}. These species are likely to form after the reaction of \ce{LiPF6} with \ce{H2O}, and later reaction with the Si NPs surface~\cite{aurbach2000,philippe2013}.
In order to explore the contribution of such species to the overall spectrum, theoretical F K-edge XAS spectra were computed for similar prototypical systems, i.e. \ce{POF3} and \ce{F3OSi}, as reported in \autoref{fig:XAS_F} (dashed lines) together with the experimental lithiated curve (red solid line) and the relative residue curve (grey area). 

These results suggest that the low energy feature in the electrode spectrum could indeed arise from transitions between $1s$ states and $\sigma^*$ orbitals in F$-$P and F$-$Si environments, similar to those found in \ce{PO_xF_y} and \ce{SiO_xF_y} species, as they match fairly well with the residue curve (grey area) in this energy region. Indeed, these contributions, though minor, could solve both the model underestimation of the intensity at low energies and the overestimation between 692 and 704 eV, which both result from the absence of reference contribution in the onset energy region (see Figure S2, Supporting Information). 

\begin{figure}
    \centering
    \includegraphics[width=0.5\textwidth]{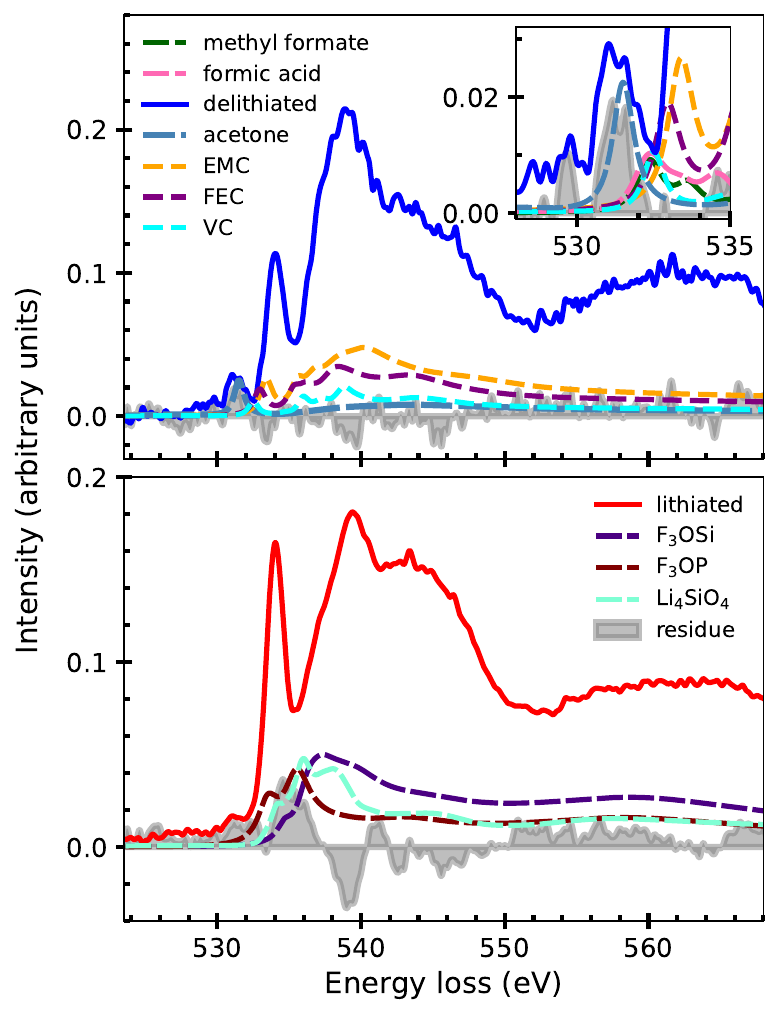}
\caption{XRS O K-edge spectrum of c-Si NPs electrodes in the low momentum transfer regime. The experimental curves in the delithiated (top, blue line) and lithiated (bottom, red line) states are compared with organic (top) and inorganic (bottom) theoretical references (dashed lines). The residual curve from ~\cref{fig:XRSall} is also reported in grey.
}\label{fig:XAS_O_sup}
\end{figure}

\emph{O K-edge.} Moving to the O K-edge (\autoref{fig:XRSall}, central panels), we observe that the XRS spectrum departs considerably from the one of the pristine electrode (green curve) already after the first charge, with a shape very much resembling that of the \ce{Li2CO3} reference (orange curve). This is confirmed by the best fit model, which identifies \ce{Li2CO3} as the main O-bearing compound in the SEI for the lithiated state (77\%). Upon delithiation, we notice an intensity decrease for the peak located at 533.9 eV, which suggests a partial decomposition of the \ce{Li2CO3} present in the SEI. This is accompanied by an increase of the contribution due to the pristine electrode reference in the linear combination model (from 18 to 43\%). Considering that the binder and the natural oxidation layer of c-Si NPs (\ce{SiO2}) are the main oxygenated species in the pristine electrode, both having similar O K-edge shape, and that the binder can be considered as fixed in the first cycles, we could possibly associate the higher pristine electrode weight with a higher oxidation level of the Si NPs (see Figure S3 and related discussion in Supporting Information). 

The main features not captured by the model ($\dagger$) are the peak at about 531 eV (both SoC, but more prominent in the delithiated state) and the underestimation (overestimation) around 534-536 eV (above 538 eV) characterizing the lithiated state. The XRS data are compared to the computed theoretical references in \autoref{fig:XAS_O_sup}, split in organic (top) and inorganic (bottom) contributions. For the organic part, we considered the simplest ketone, ester and carboxyl-based compounds, to try to associate the spectral contributions to specific local chemical environments. It is clear from the inset of the top panel that the contribution of the organic species to the spectrum should be minor and localized in the 530-533 eV region, with the ketone group accounting for the feature at about 531 eV. 

The inorganic compounds used as references are \ce{POF3} and \ce{F3OSi}, as resulting from the F K-edge analysis, and \ce{Li4SiO4}, as a prototype of Li-silicates expected to form upon lithiation of the \ce{SiO2} natural layer of Si NPs \cite{philippe2013}. 
Their contribution to the overall spectrum is distributed in a wider energy range than that of the organic ones, mostly overlapping with the underestimation/overestimation of the model in the lithiated state described above. In particular, the depth change for the dip at 535 eV could be associated to the formation of lithium silicates and \ce{PO_xF_y} species. Overall, the contribution of other inorganic compounds remains relatively small when compared to \ce{Li2CO3}, which confirms the picture drawn from the F K-edge analysis.


\emph{C K-edge.} Finally, the C K-edge (\cref{fig:XRSall}, left top panel) reveals the formation of the carbonaceous portion of the SEI during the initial charge and its evolution upon the first discharge. For both SOC, the linear combination analysis indicates the presence of a variable mixture of pristine electrode (Na-CMC+SP) and \ce{Li2CO3}, identifying the latter as the principal C-bearing compound in the SEI. The accuracy of the fitting indicates only a minor missing contribution in the 285-290 eV energy region, which can be attributed to organic species from the electrolyte according to DFT simulations, as detailed in Supporting Information (see Figure S4). Notably, upon delithiation, the quantity of \ce{Li2CO3} is reduced, confirming its partial decomposition, as indicated by the analysis of the O K-edge. This finding agrees with previous observations by Philippe et al.~\cite{philippe2013}, which reported variable-thickness SEI by monitoring the C $1s$ peak by soft and hard XPS during the first cycle. 

\subsection{Si edge: Si-NPs evolution and \ce{Li_xSi} phases}\label{subsec:silicon}

\begin{figure}[b!]
    \centering
    \includegraphics[width=0.5\textwidth]{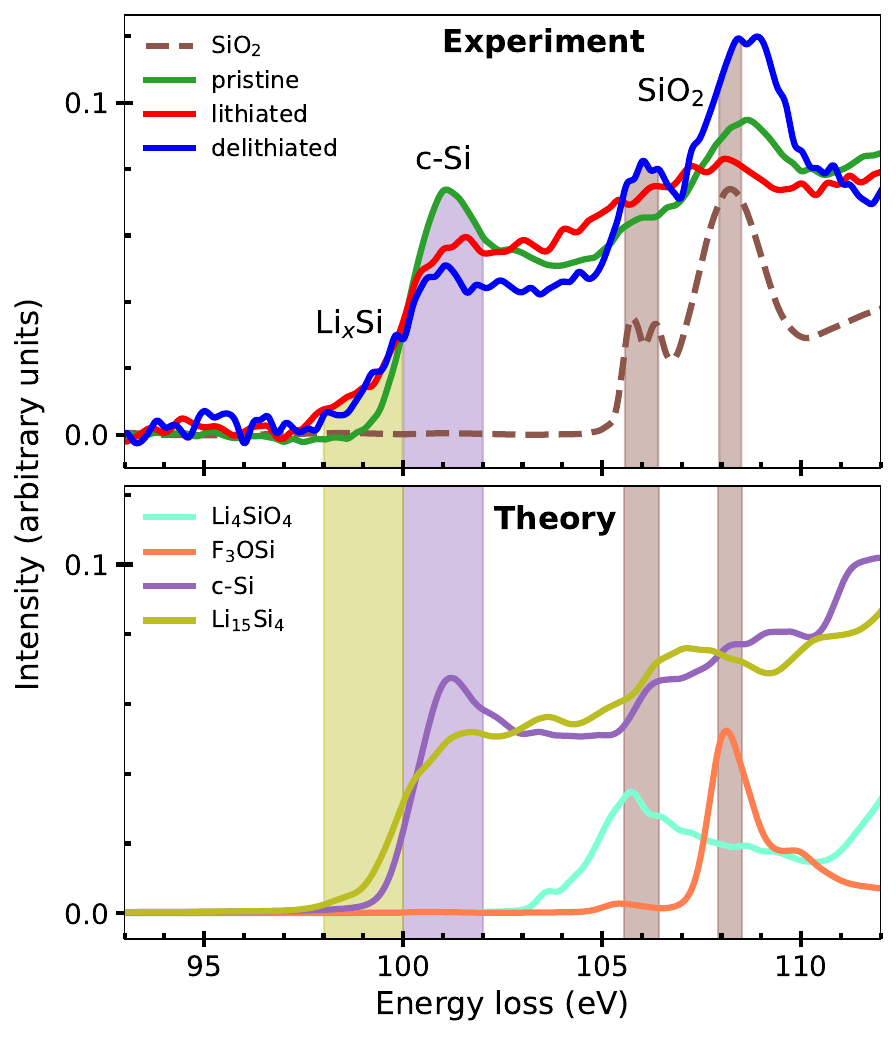}
\caption{XRS Si \ce{L_{2,3}} edge spectrum of c-Si NPs electrodes in the low momentum transfer regime (top panel) compared with experimental \ce{SiO2}\cite{turishchev2019} (dashed brown lines) and theoretical references (bottom panel).}\label{fig:XRS_Si_Ledge}
\end{figure}

After discussing the SEI formation and evolution during the first charge and discharge, we next analyse the evolution of c-Si NPs by tracking the changes with respect to the pristine signal on the Si \ce{L_{2,3}} edge.
\autoref{fig:XRS_Si_Ledge} (top) shows the Si \ce{L_{2,3}} edge spectra of the electrode in pristine (green), lithiated (red), and delithiated (blue) conditions. In order to support the analysis, we include an experimental reference for \ce{SiO2}~\cite{turishchev2019} and a group of theoretical references (bottom panel) for c-Si, \ce{Li15Si4}, \ce{F3OSi} and \ce{Li4SiO4}. The spectrum for the pristine electrode (green) indicates primarily the presence of c-Si ($\approx$94\%), with a minor amount of \ce{SiO2} ($\approx$6\%) created during the electrode preparation (detailed quantification can be found in Supporting Information, see Figure S5). This maps directly to the NP morphology, which is composed by a crystalline core surrounded by a natural \ce{SiO2} shell, as identified by the peak at 101 and by the features at 106 and 108.3 eV, respectively. 

Upon charge, the features corresponding to c-Si and \ce{SiO2} are severely dampened, with a corresponding increase of the signal below 100 eV and between 102 and 107 eV. The very same modifications are observed when comparing the computed spectrum for c-Si and a prototypical Li-silicide, i.e. \ce{Li15Si4}. This alloy corresponds to the full lithiation state usually obtained in standard coin cell cycling, with a lithiation index x=3.75.  
We observe a close resemblance of the computed \ce{Li15Si4} to the experimental lithiated spectrum, suggesting the full lithiation of a major part of the NP volume, leaving only a relatively small fraction of c-Si. Indeed, the lithiation of Si NPs is usually characterized by the coexistence of a crystalline core with an outer shell highly lithiated and amorphized~\cite{tardif2017}. Analysing the data using a combination of c-Si and \ce{Li15Si4} spectra shows an excellent fitting of the spectra at 100\% SoC, yielding 9\% of remaining cristalline silicon, as seen in Figure S6.

Upon discharge, the electrode spectrum features an attenuation of the peak at 101 eV, which could be attributed to the partial amorphization of the c-Si NPs after lithiation and delithiation, with the resulting spectrum containing characteristics of both crystalline and amorphous Si phases. Moreover, we observe the persistence of the lower onset energy (below 100 eV) found for the lithiated state, which suggests that a small quantity of Li is trapped in the NP in the form of \ce{Li_xSi}. Finally, the peaks located at 106 and 108.3 eV display a higher intensity as compared to the spectrum of the pristine sample. This may be related to an increased amount of oxidized Si, as already evidenced on the O K-edge. This could be possibly produced by a native oxide regrowth~\cite{schroder2012}, as we cannot exclude a short oxygen exposure in the glovebox from the sample preparation. 
Minor contributions from \ce{SiO_xF_y} and \ce{Li_xSiO_y} species can also not be discarded since their main features 
overlap with the peak structure attributed to \ce{SiO2}, according to DFT simulations. Attempts to fit the data using a combination of amorphous/cristalline silicon and native oxide components clearly fail (Figure S6), proving the presence of a significant amount of remaining \ce{Li15Si4} phase (typically around 20-25\%, see Figure S6). Moreover, it is seen that the excess intensity below 100eV remains, sustaining the presence of other (less) lithiated species. In fact, during delithiation, phases with lithiation index less than 3.75 are formed, and not-fully delithiated regions might contain a distribution of poorly lithiated particles, tipically with x $<$ 2. 




\subsection{Li K-edge: analysing changes in lithium environments}\label{subsec:li-k edge}

\begin{figure}
    \centering
    \includegraphics[width=0.5\textwidth]{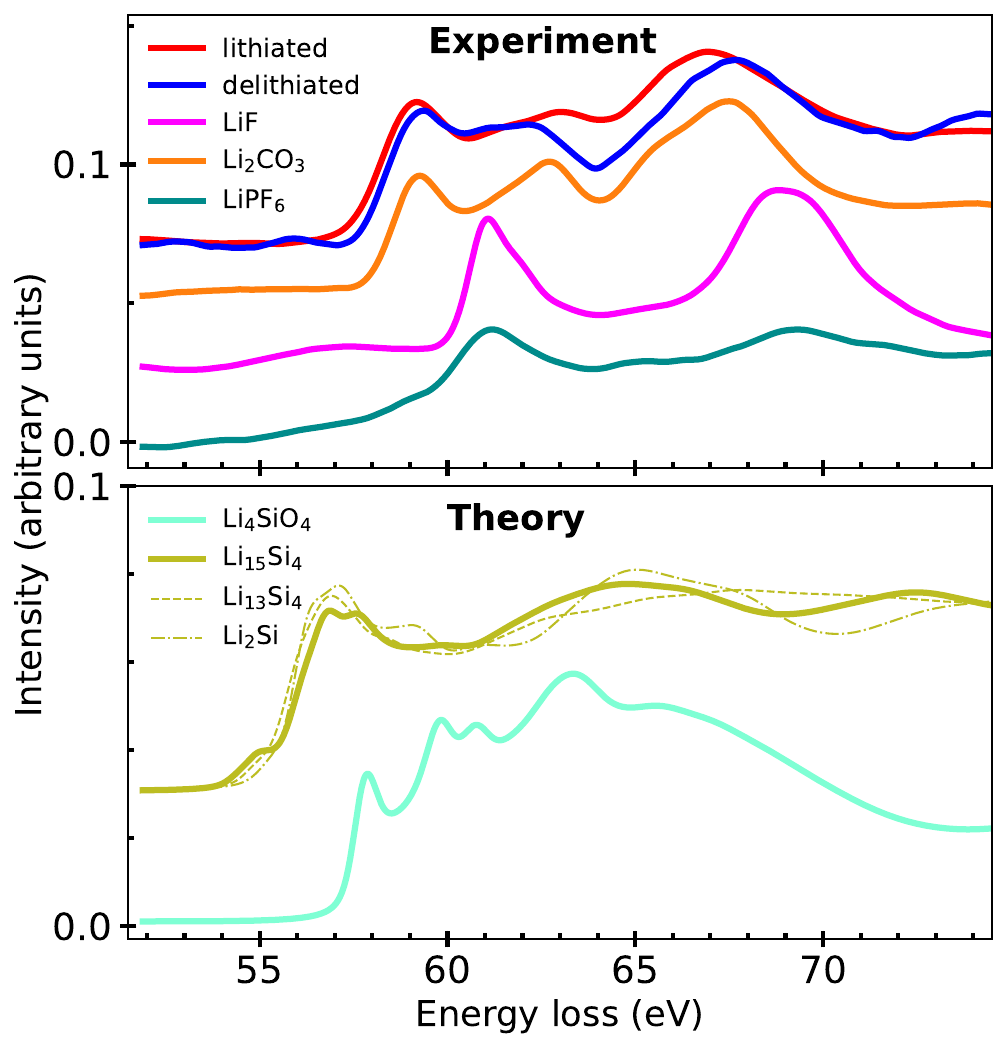}
\caption{XRS Li K-edge spectrum of c-Si NPs electrodes in the high momentum transfer regime (top) compared with experimental (top) and theoretical references (bottom).}\label{fig:Li_K_edge}
\end{figure} 

To conclude the qualitative analysis, the Li K-edge spectra of c-Si NP electrodes in the high momentum transfer regime are shown in \autoref{fig:Li_K_edge} (top) for both lithiated (red) and delithiated
(blue) conditions. These curves are compared with experimental references for \ce{Li2CO3}, \ce{LiF} and \ce{LiPF6} (top), as well as theoretical references for various \ce{Li_xSi} phases and \ce{Li4SiO4} (bottom). Qualitatively, we find that after the initial lithiation, the electrode very much resembles \ce{Li2CO3}, with relatively minor amounts of \ce{LiF} and \ce{LiPF6} localized both at 61 eV and 69 eV. 
However, the LC model based on these three compounds cannot reproduce neither the lower energy onset nor the increased intensity between 62 and 67.5 eV with respect to \ce{Li2CO3} (see Supporting Information Figure S7). The comparison with the calculated spectra of \ce{Li_xSi} and \ce{Li4SiO4} (bottom panel) suggests that both these features could be attributed to Li-silicides (and silicates), which indeed show lower energy onset and higher intensity in the 62-67 eV region. 
Furthermore, these features and the ones corresponding to \ce{Li2CO3} undergo significant changes after discharge, when \ce{Li^+} is extracted from the NPs, thus confirming the observation from the other edges of a partial \ce{Li2CO3} decomposition upon delithiation. 

To further evaluate the proportions of the different lithium bonding environments, we fitted the experimental data using a combination of experimental and numerical XRS spectra (Figure S7). At 100\% SoC, the Li edge is remarkably well reproduced using 50\% of \ce{Li2CO3} and 47\% of \ce{Li15Si4}. 
Also, we notice that the LiF and \ce{LiPF6} compounds are not strictly needed to nicely reproduce the Li edge spectra (see comments in Supporting Information, Figure S7), which shows that amounts of these compounds remain relatively low with respect to carbonates and Silicon phases. The 0\% SoC Li edge data are more difficult to reproduce, as already seen for other edges analysis. Fitting the data using the combination of 
\ce{Li15Si4},\ce{Li2CO3},\ce{LiF}, and \ce{LiPF6} spectra enables to reproduce only ca. 95\% of the delithiated electrode spectra. Even by injecting the knowledge gained using F edge, fixing the ratio of 
fluorine species, we cannot fully describe, particularly, the region at 59-60 eV (Figure S7). 

As recently shown in the literature, analysing lithium chemical environment using XRS is particularly challenging as the peaks in this region are containing overlapped features of \ce{Li_xSiO_y} and \ce{Li_xSi}. In fact, our Si L$_{2,3}$ edge analysis showed the presence of such distribution of lithiated states for oxides and silicon. Nevertheless, we understand from the changes observed in our XRS data from SoC 100\% to 0\% that chemical heterogeneities are increased during the delithiation steps, which originates from both the non-uniform delithiation of the silicon, as well as the partial dissolution/recombination of SEI products.

%

\subsection{Origin of Li loss}\label{subsec:li-loss}

The XRS analysis revealed the presence of several lithium environments that build on the first lithiation and evolve during successive delithiation, showing, particularly, the partial dissolution of the SEI and trapping of lithiated silicides. To further classify and quantify this Li loss, we discuss the correlations between XRS findings and electrochemical data. First we express the total capacity after the first lithiation $Q^{lit}$ (red dot in \autoref{fig:galvanostatic}) as
\begin{equation}\label{Q_lit}
Q^{lit} = Q^{lit}_{SEI_I} + Q^{lit}_{Li_xSi} + Q^{lit}_{SEI_{II}},
\end{equation}
where $Q^{lit}_{SEI_I}$, $Q^{lit}_{Li_xSi}$ and $Q^{lit}_{SEI_{II}}$ are the portions of the total capacity involved in the formation of fluorinated SEI, \ce{Li_{x}Si} and \ce{Li2CO3}-rich SEI, respectively. Similarly, following delithiation the retrieved capacity $Q^{delit}$ (blue dot in \autoref{fig:galvanostatic}) is
\begin{equation}\label{Q_delit}
Q^{delit} = Q^{delit}_{Li_xSi} - Q^{delit}_{SEI_I}- Q^{delit}_{SEI_{II}},
\end{equation}
being $Q^{delit}_{Li_xSi}$($Q^{delit}_{SEI_I}+Q^{delit}_{SEI_{II}}$) the capacity extracted (lost) from the Si NPs (SEI evolution) upon delithiation. After combining \autoref{Q_lit}, \autoref{Q_delit} and defining the capacity loss due to the Li trapped inside the NPs as $Q_{Li_{trap}}=Q^{lit}_{Li_xSi}-Q^{delit}_{Li_xSi}$, we can express the total capacity loss as 
\begin{equation}\label{Q_loss}
Q^{lit}-Q^{delit} = Q^{lit}_{SEI_I}+ Q^{delit}_{SEI_I}+ Q^{lit}_{SEI_{II}}+Q^{delit}_{SEI_{II}}+ Q_{Li_{trap}}.
\end{equation}

Under the previous assumptions, $Q^{lit}$, $Q^{delit}$ and $Q^{lit}_{SEI_I}$ can be directly retrieved from \autoref{fig:galvanostatic}. Concerning the amount of Li trapped in Li–Si alloys, recent titration-gas chromatography experiments conducted on similar Si NPs with a Na-CMC binder~\cite{sreenarayanan2022} have found $Q_{Li_{trap}} \sim  19\%$ of the total capacity loss after the first cycle. We extrapolate their method to our case to estimate the quantity of trapped lithium, that we find close to 17\% (details are given in Supporting Information, Figure S8). With this assumption, we find that the total capacity loss, quantified at 547  mAhg$^{-1}$, is divided into 17\% due to the residual \ce{Li_{x}Si} and 83\% to the processes involving the formation and decomposition of the SEI.

Based on this hypothesis, we now use XRS data to further understand the 83\% lost in the SEI, using the most reliable and complete edge, e.g. C-edge. Indeed, the percentage of \ce{Li2CO3} decomposition upon delithiation can be quantified by correlating the linear combination weights of Na-CMC+SP and \ce{Li2CO3} obtained for the C K-edge with the relative quantities of carbon atoms. If we define $\ce{C^i_{X}}$, $\ce{w^i_{X}}$ and $\ce{m^i_{X}}$ as the amount of C atoms, the linear combination weight, and the mass, respectively, corresponding to the compound $X$ in the SOC $i=d,l$ (where $d$ stands for delithiated and $l$ for lithiated), then we can establish that \(\frac{\ce{C^i_{Na-CMC+SP}}}{\ce{C^i_{\ce{Li2CO3}}}} \equiv \frac{\ce{w^i_{Na-CMC+SP}}}{\ce{w^i_{\ce{Li2CO3}}}}\). 
Assuming a negligible decomposition of the binder and the conductive additive (i.e., \ce{C^d_{Na-CMC+SP}}=\ce{C^l_{Na-CMC+SP}}), we obtain \( \frac{\ce{w^d_{Na-CMC+SP}}}{\ce{w^d_{\ce{Li2CO3}}}}\ce{C^d_{\ce{Li2CO3}}} = \frac{\ce{w^l_{Na-CMC+SP}}}{\ce{w^l_{\ce{Li2CO3}}}}\ce{C^l_{\ce{Li2CO3}}}\), leading to \( \frac{\ce{C^d_{\ce{Li2CO3}}}}{\ce{C^l_{\ce{Li2CO3}}}} = \frac{\ce{m^d_{Li2CO3}}}{\ce{m^l_{Li2CO3}}}=0.65\). 
This corresponds to a 35\% of decomposition of the \ce{Li2CO3} formed during the first lithiation. 
These considerations allow us to express the total capacity loss of \autoref{Q_loss} as:
\begin{equation}\label{Q_loss_final}
Q^{lit}-Q^{delit} = Q^{lit}_{SEI_{I}}+1.35Q^{lit}_{SEI_{II}}+  Q_{Li_{trap}}
\end{equation}
Using the results from \autoref{fig:galvanostatic} and \autoref{Q_loss_final}, we can approximately split the total capacity loss during the initial cycle into 17 \% from the remaining \ce{Li_{x}Si}, 30 \% from the formation of F-rich SEI, and 53 \% from the \ce{Li2CO3}-rich SEI part. As said earlier, it is difficult to cross-check these quantities using Si and Li edge data as they are not fully described using our set of reference and simulated spectra. Nevertheless, the 100\% SoC data provide some additional indications. From the Li edge, we know that ca. half of the signal is arising from Li in \ce{Li_15Si_4}, while the rest is mostly \ce{Li_2CO_3}. A simple composition-based calculation shows that for 100 silicon atoms, 375 lithium ions would be expected to be inserted at full lithiation, and deinserted during delithiation. The irreversible capacity corresponds to a loss of 60 ions, with 42 lithium trapped in \ce{Li_15Si_4}. This would correspond to 12 Silicon. Hence, the percentage of silicon disconnected after the first cycle and stuck in partially lithiated states, would be around 10\%. At that point, the Li edge data indicate that for one lithium trapped in Silicon, there is 0.13 lithium immobilized in the SEI, a ratio that seem compatible with observed SEI thicknesses on nanoparticles which are in the range of several nanometers for particles below 100-nm diameter \cite{kumar2020}.


\section{Conclusions}

The XRS spectra of anodes composed of crystalline Si NPs, probed at different states of charge, have unique chemical signatures demonstrating both reversible and irreversible changes in the electrode materials on cycling. These signatures can be differentiated when comparing qualitatively the high-resolution real electrode spectra acquired in the energy range of light elements (C, O, Li, F, Si, Li) with well-chosen representative reference compounds as \ce{LiPF6}, LiF, \ce{Li2CO3}, carbon binder phase, silicon powder, and pristine silicon electrode. 

A semi-quantitative analysis performed by linear decomposition using experimental spectra for C, O and F edges, aided by comparison with DFT-based data, evidences the presence of \ce{Li2CO3}, \ce{LiPF6} and LiF after one cycle, and potentially minor quantities of additional degradation products, such as  \ce{PO_xF_y} and \ce{SiO_xF}.
Weight factors of relevant reference compounds are extracted, revealing how they are formed or dissolved during (de)lithiation. Specifically, we find an evident decrease in the peak intensities associated with carbonates when the electrode is delithiated,
while the ratio of \ce{LiPF6} and LiF amounts appear to remain nearly stable. The SEI, which grows during lithiation due to the decomposition of organic electrolytes (EMC and FEC), additives (VC), and salt (\ce{LiPF6}), is thus found to partially dissolve upon delithiation, in agreement with previous  studies~\cite{philippe2013}. 


Simulations also provide insights to interpret Si and Li edges and discuss correlations between the amounts of lithium trapped in the SEI and in the silicon. In particular, the SEI-breathing is accompanied by an incomplete delithiation, which relates to the significant capacity loss in the first cycle. Indeed, we evidence that lithium is not fully removed from the anode in the delithiated state, even after only one cycle. The irreversible lithium loss due to trapping in alloyed Si regions is quantified at 2.7\% of Li unavailable for reactions, in agreement with other studies \cite{sreenarayanan2022}. 

XRS is suited to detect changes in the chemical environment of the silicon active particles as well as their degradation processes, providing insights into the degree of amorphization of silicon and presence of silicon oxides, the amounts of carbonates and fluorine-based species and their dynamics during the lithiation/delithiation sequence, as well as, potentially, revealing the presence of a distribution of remaining lithiated phases in a cycled material. This is potentially extremely interesting for investigating aging and composite materials, where the long-term cycling effects could be quantified as well as the impact of sensitive parameters as silicon content, size, coatings, etc. 

As a consequence, while further investigations are surely needed to better comprehend these phenomena, we here anticipate that XRS can be successfully applied to investigate in a semi-quantitative way a wide range of battery materials, enabling to gather high-fidelity bulk information, highly complementary to invasive surface-limited XPS. Spectra simulations based on ab initio DFT approaches also prove valuable to interpret and complement XRS measurements, and -- validated towards more extended sets of data -- could be critical to predict SEI features in a variety of materials for unexplored post-lithium ion technologies.

\section{Materials and methods}

\subsection{Sample preparation} 

Crystalline SiNPs of size between 60 and 100 nm were synthesized by laser pyrolysis from high purity silane~\cite{desrues2019}. Electrodes were prepared by mixing  50 wt\% of SiNPs, 25 wt\% of sodium carboxymethyl cellulose (Na-CMC, Merck), and 25 wt\% carbon black (Super P). The powders were ground and dissolved in purified water (18.2 $\Omega$ at 25 $^{\circ}$C) to produce a slurry, which was deposited on a Cu foil (20 $\mu$m, thickness) and dried for 12 h at 80 $^{\circ}$C.

Coin cells were prepared by cutting out circular disks of this Si-based material (0.97 mg/cm$^2$), assembled with carbonate-mixture based electrolyte, celgard 2400 (monolayer polypropylene) separator, and lithium metal. The electrolyte used was composed by 1 M \ce{LiPF6} (3FEC/7EMC,v/v) with 2 wt\% vinyl carbonate (VC). 
The cells were cycled and stopped at a given potential for obtaining different states of charge (SoC) conditions. Subsequently, each coin cell was disassembled and transferred without rinsing to designed airtight cell in a glovebox with Ar to probe Li, Si, C, O and F chemical environments by post-mortem XRS measurements. 

\subsection{XRS measurements} 

XRS experiments were performed on the beamline ID 20 of the European Synchrotron Radiation Facility (ESRF, Grenoble, France). The pink beam from four U26 ondulators was monochromatized to an incident energy of 9.6837 keV, using a cryogenically cooled Si(111) monochromator and focused to a spot size of approximately 10 mm $\times$ 20 mm (V $\times$ H) at the sample position using a mirror system in Kirkpatrick–Baez geometry. The large solid angle spectrometer at ID20 was used to collect XRS data with 72 spherically bent Si(660) analyzer crystals. The data were treated with the XRStools program package as described elsewhere~\cite{sahle2015,huotari2017}.

The powder samples were pressed into a pellet, which was placed into the beam so to have a 10 grazing incident beam. All XRS measurements were collected at room temperature. Full range scans were collected from 0 to 700 eV with a 1 eV step size. After the acquisition of the broad scan, several detailed scans of specific edges were collected with a 0.2 eV step by scanning the incident beam energy to record energy losses in the vicinity of the Li K-edge (54.7 eV), C K-edge (284.2), O K-edge (525 eV), F K-edge (696.7) and Si L-edge.

Acquisition scans lasted around 2–4 h per sample. All scans were checked for consistency before averaging over them. The overall energy resolution of the XRS spectra was 0.7 eV, as estimated from the FWHM of elastic scattering from a piece of adhesive tape. Also, signals from analyzer crystals at different scattering angles were measured, covering a momentum transfer from 2.5 to 9.2 \AA$^{-1}$. The data were integrated at high q for further analysis.

\subsection{Computational details}
XRS spectra in the low momentum transfer regime were computed with the XSpectra~\cite{gougoussis2009,gougoussis2009a,bunau2013} code of Quantum ESPRESSO~\cite{giannozzi2009,giannozzi2017,giannozzi2020}, while for the high momentum transfer XRS we used a modified version of XSpectra~\cite{gallerande2018}. Crystalline and molecular structures used as theoretical references were obtained from the Materials Project~\cite{jain2020} and PubChem~\cite{kim2022} databases respectively. Ultrasoft pseudopotentials from the SSSP library v1.1~\cite{prandini2018} with PBE exchange correlation functional were used for ground state calculations and non-absorbing atoms. Core-hole pseudopotentials were instead generated to describe excited atoms for the simulation of K-edge (C, O, F and Li) and \ce{L2,3} edge (Si) spectra. The energy cutoffs for the kinetic energy of the wave functions and the charge density were set to 80 Ry and 640 Ry, respectively. A supercell approach was employed to minimize interactions between excited atoms and their periodic images, maintaining a minimum distance of 8 {\AA} between replicas in crystalline systems. In the case of molecules, the supercell size was chosen according to requirements of the Martyna-Tuckerman~\cite{martyna1999} method for treating the system as isolated. The Brillouin zone corresponding to the supercell was sampled at the $\Gamma$ point for molecules and using a Monkhorst-Pack mesh with a k-points grid spacing of 0.25~\AA{}$^{-1}$ in the case of crystals. In addition, all the spectral calculations were conducted with a 4x4x4 k-point grid with an energy-dependent broadening parameter~\cite{gallerande2018}. We applied different core-hole treatments depending on the system, i.e. excited core-hole (XCH)~\cite{prendergast2006} for crystals and the transition-potential approach (half core-hole HCH)~\cite{triguero1998} or the full-core hole (FCH) for molecules. Each spectrum was aligned with respect to the excitation onset, expressed as the binding energy of the edge core state, computed within the Delta Kohn-Sham ($\Delta$KS) \cite{ljungberg2011,walter2016} approach and referred to the highest occupied state in the presence of the core-hole \cite{england2011,guo2023}.

\section*{Author contribution statement}
SL, DP, SP and EM conceived and supervised the work. NH synthesized the silicon nanoparticles. DZD and CB prepared the electrodes. ST and CB designed the sample holder. SL, SP, ST, DZD and CB performed the XRS experiments, aided by CS and CC. MHB computed the numerical spectra, with the support of IM and under the supervision of DP and EM. MHB correlated the experimental and numerical results, aided by DZD, AL, SL and DP. MHB and DZD wrote the first draft, together with SL and DP. All authors discussed the results and commented on the manuscript at all stages. 

\section*{Declaration of competing interest}
The authors declare that they have no known competing financial interests or personal relationships that could have appeared to influence the work reported in this paper.

\begin{acknowledgement}
This work was supported by the European Union’s Horizon 2020 research and innovation programme (BIG-MAP, Grant No. 957189, also part of the BATTERY 2030+ initiative, Grant No. 957213), by the European High Performance Computing Joint Undertaking (MaX Centre of Excellence -- Materials design at the eXascale, program HORIZON-EUROHPC-JU-2021-COE01, Grant No. 101093374), and by the European Union -- NextGenerationEU, through the MUR -- Prin 2022 programme (2D-FRONTIERS, Grant No. 20228879FT). The authors acknowledge the CINECA award under the ISCRA initiative, for the availability of high-performance computing resources and support. The authors acknowledge the European Synchrotron Radiation Facility (ESRF) for providing beamtime (proposal IH-MA-4). Data analysis was discussed in the frame
of the Grenoble Battery Pilot Hub, proposal MA4929: “Multi-scale multitechniques investigations of Li-ion batteries: towards a European Battery Hub.”
\end{acknowledgement}

\begin{suppinfo}

Details on the LC fitting model and on the analisys of multi-edge XRS spectra. 

\end{suppinfo}

\providecommand{\latin}[1]{#1}
\makeatletter
\providecommand{\doi}
  {\begingroup\let\do\@makeother\dospecials
  \catcode`\{=1 \catcode`\}=2 \doi@aux}
\providecommand{\doi@aux}[1]{\endgroup\texttt{#1}}
\makeatother
\providecommand*\mcitethebibliography{\thebibliography}
\csname @ifundefined\endcsname{endmcitethebibliography}
  {\let\endmcitethebibliography\endthebibliography}{}

\newpage


\section*{Supplementary material}

# SUPPORTING INFORMATION

# Understanding the irreversible lithium loss in silicon anodes using multi-edge X-ray scattering analysis

Michael A. Hernandez Bertran, Diana Zapata Dominguez, Christopher Berhaut, Samuel Tardif, Alessandro Longo, Christoph Sahle, Chiara Cavallari, Ivan Marri, Nathalie Herlin-Boime, Elisa Molinari, Stéphanie Pouget, Deborah Prezzi,* and Sandrine Lyonnard*

E-mail: deborah.prezzi@nano.cnr.it; sandrine.lyonnard@cea.fr

## Multi-edge XRS data

Figure S1 reports, in the top panels, the XRS spectra for the fully lithiated (red) and delithiated (blue) electrodes, measured on the C, O, F and Li K-egde and on the Si L-edge. The difference between lithiated and delithiated SoC spectra is also displayed (grey area) to highlight the spectral ranges where the two SoC differ the most. The bottom panels show the XRS spectra for a set of reference compounds for each edge, including both pristine electrode species and known degradation products, measured in the same conditions and set-up.

1

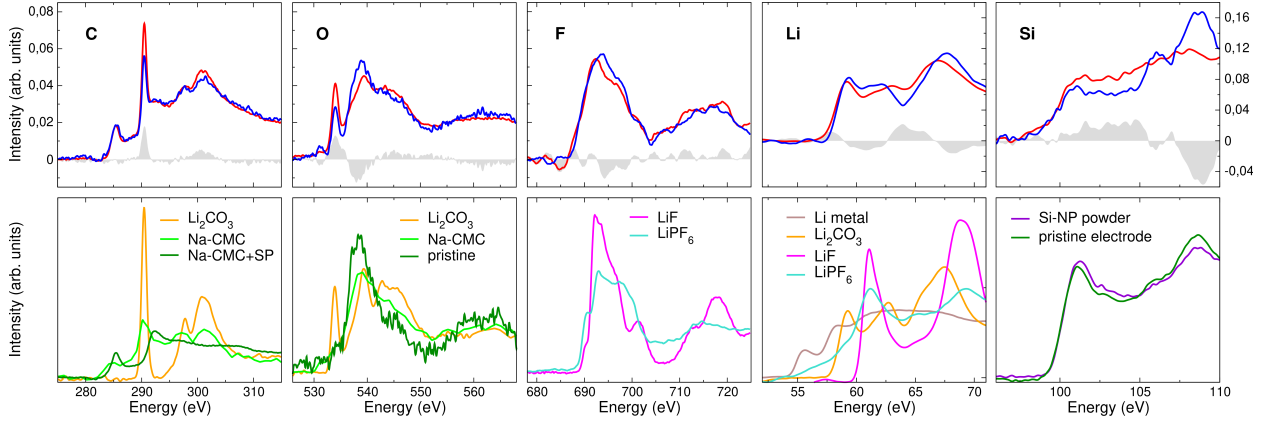

Figure S1: (Top panels) Experimental non-resonant inelastic X-ray scattering spectra of c-Si NPs electrodes for the C, O, F and Li K-edges and the Si $L_{2,3}$-edge in the lithiated (red) and delithiated (blue) SoC, as well as their difference (grey area). (Bottom panels) XRS spectra for the reference compounds.

# Fitting procedure

To quantify the correlations between the spectra of lithiated/delithiated electrodes and those of the reference compounds, a linear combination fitting of XRS data with the reference spectra is performed. For the linear combination (LC) procedure, we express the target spectrum $|I\rangle$ in terms of a basis of known references $|I_i\rangle$ as

$$|I\rangle = \sum_{i=1}^{n} c_i |I_i\rangle, \qquad (1)$$

where the vectors $|I\rangle$ and $|I_i\rangle$ have $m$ components determined by the number of energy bins of the experimental data. Since the size of the basis is smaller than the number of experimental points ($n < m$) the coefficients $c_i$ are found approximately using the least squares method as follows

$$c_j = \sum_{i=1}^{n} ||\langle I_i | I_j \rangle||^{-1} \langle I_i | I \rangle. \qquad (2)$$

For all the cases we considered that the target spectrum $|I\rangle$ and all the references $|I_i\rangle$ are normalized in the same energy range. Furthermore, a LC is considered valid only if all

the coefficients are non negative ($c_i \geq 0$).

# Fitting for the F K-edge

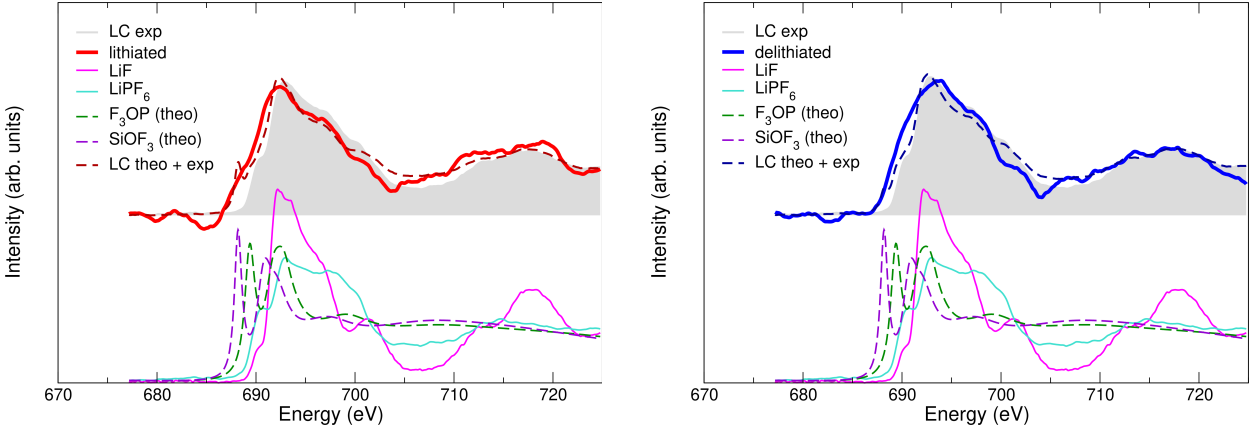

Figure S2: XRS F K-edge spectrum for the lithiated (left panel, red solid line) and delithiated (right panel, blue solid line) electrodes are compared with the LC fitting obtained using either the experimental reference data only (grey area) or both theory and experimental reference data (dark red/blue dashed lines). The coefficients of the LC are (LiF, $LiPF_6)_{exp}$ = (0.29, 0.65) / (0.31, 0.66) and (LiF, $LiPF_6$, $F_3OP$, $F_3OSi)_{theo+exp}$ = (0.31, 0.22, 0.17, 0.31) / (0.29, 0.41, 0.19, 0.13) for the lithiated / delithiated sample, respectively. The spectra for the reference compound are also reported, with some y-displacement to ease readability.

Figure S2 shows the XRS F K-edge spectrum for the lithiated (left panel, red solid line) and delithiated (right panel, blue solid line) electrodes and the comparison with the LC fitting curves. The grey area displays the fitting curve obtained by using the experimental references only, i.e. LiF and $LiPF_6$. The dashed dark red and dark blue curves show instead the LC fitting obtained by including both theory and experimental references. The latter is just meant to be a guide for the eyes, as we only considered a couple of prototypical environments without presumption of completeness (in addition to all the accuracy issues connected to the simulations themselves in view of the various approximations adopted). With all these caveats, we indeed see an improvement in the fitting for the lithiated case. For the delithiated case, despite allowing to fill in the onset region, the inclusion of these additional references fails in reproducing the shift to higher energies of the edge maximum,

which is also the main difference between the two SoC when we compare the two curves in Figure S1. This suggests some evolution of the F-bearing compounds during delithiation.

## Variations on the O K-edge

As mentioned in the main text, the main difference between the two SoC as resulting from the linear combination model for the O K-edge is the decrease of $Li_2CO_3$ contribution accompanied by an increase of the contribution due to the pristine electrode. In order to better understand this change, we need to refer to some constant quantity as these percentages are relative to the total amount of O at the end of the lithiation/delithiation process during the first cycle. To do so, we first focus on C K-edge, where we can assume that the binder (Na-CMC) is constant during the whole first cycle. With this assumption (66% Na-CMC+SP in the lithiated SoC $\approx$ 74% Na-CMC+SP in the delithiated SoC), we can renormalize the percentage of $Li_2CO_3$ in the lithiated SoC (wrt the delithiated basis), which gives 37% of $Li_2CO_3$ in the lithiated wrt the delithiated SoC. This implies a ratio between lithiated and delithiated $Li_2CO_3$ of 37/24=1.54. If we perform the same analysis on the O K-edge, assuming the pristine signal as constant, we obtain a ratio between the $Li_2CO_3$ in the two SoC of 3.35. This means that the assumption of a constant pristine electrode does not hold. The only explanation is that we are fitting with the pristine signal the increment of $SiO_2$ occurring upon delithiation. This is possible due to the similarity of the O K-edge of Na-CMC and $SiO_2$, as shown in Figure S3.

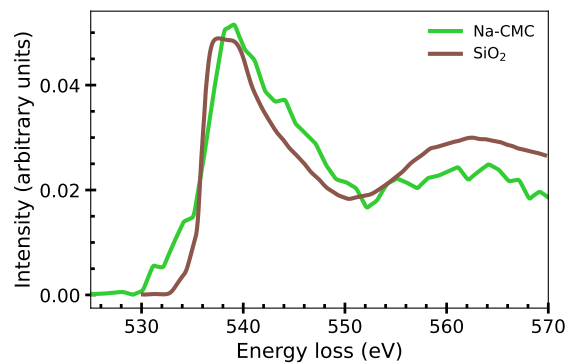

Figure S3: XRS O K-edge spectrum of Na-CMC in the low momentum transfer regime, as compared to $SiO_2$ XANES from Ref. 1.

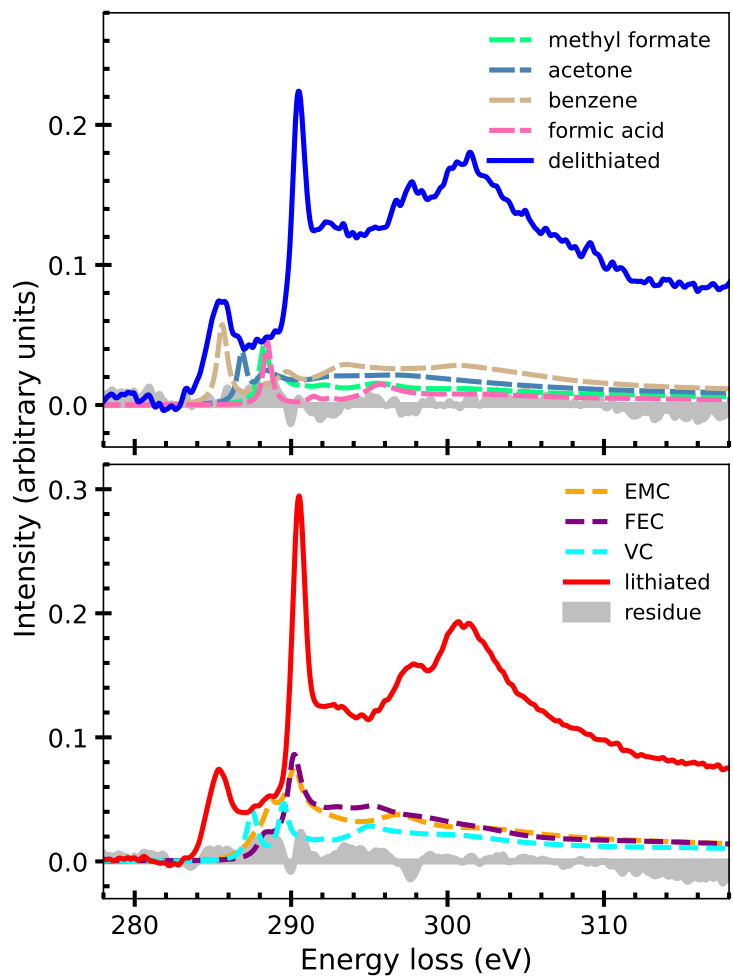

Figure S4: XRS C K-edge spectrum of c-Si NPs electrodes in the low momentum transfer regime. The experimental curves in the delithiated (top, blue line) and lithiated (bottom, red line) states are compared with organic theoretical references (dashed lines). The residual curve from Figure 2 of the main text is also reported in grey.

5

# Computational C K-edge spectra for organic groups of relevance

## Quantification of SiO$_2$ content

Figure S5 displays the XRS Si L$_{2,3}$-edge spectrum of the Si powder used in electrode preparation (violet curve), the pristine c-Si NPs electrode (green curve), as well as a reference spectrum for SiO$_2$, as taken from Ref. 2 (dashed brown curve). In order to quantify the amount of c-Si reverting to SiO$_2$ during the preparation procedure, we consider the Si$_{\text{pristine}}$ area, which accounts for the number of transitions in the energy range up to the first feature of SiO$_2$ for the pristine electrode, and the $\Delta$Si area, which corresponds to the difference in number of transitions in the energy range up to the first feature of SiO$_2$ between the Si powder and the pristine electrode. The area ratio $\frac{\Delta Si}{\Delta Si + Si_{pristine}}$ quantifies the relative amount of c-Si that is oxidized during the preparation process.

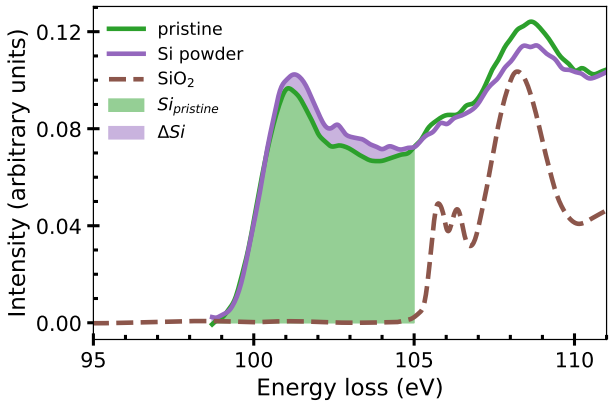

Figure S5: XRS Si L$_{2,3}$-edge spectrum of pristine c-Si NPs electrode, Si powder used in electrode preparation and SiO$_2$ from Ref. 2 (dashed lines).

It is also worth noting that, for this quantification, it is worth referring to the Si L$_{2,3}$ edge, where the contribution from c-Si and SiO$_2$ are neatly separated. Any further consideration of this sort on other edges has to be performed with care, as spectral contributions from different compounds are more overlapping and/or very similar. This is the case of the O

6

K-edge, as exemplified in Figure S3.

## LC fitting for the Si $L_{2,3}$ edge

Figure S6 (left panel) shows the comparison of the lithiated (top) and delithiated state with given references, that is, $Li_{15}Si_4$ (theory, top) and c- and a-Si (experiment, bottom). The right panel reports the LC fitting model using both experimental (c-Si, a-Si, $SiO_2$) and theoretical references ($Li_{15}Si_4$, $F_3OSi$ and $Li_4SiO_4$).

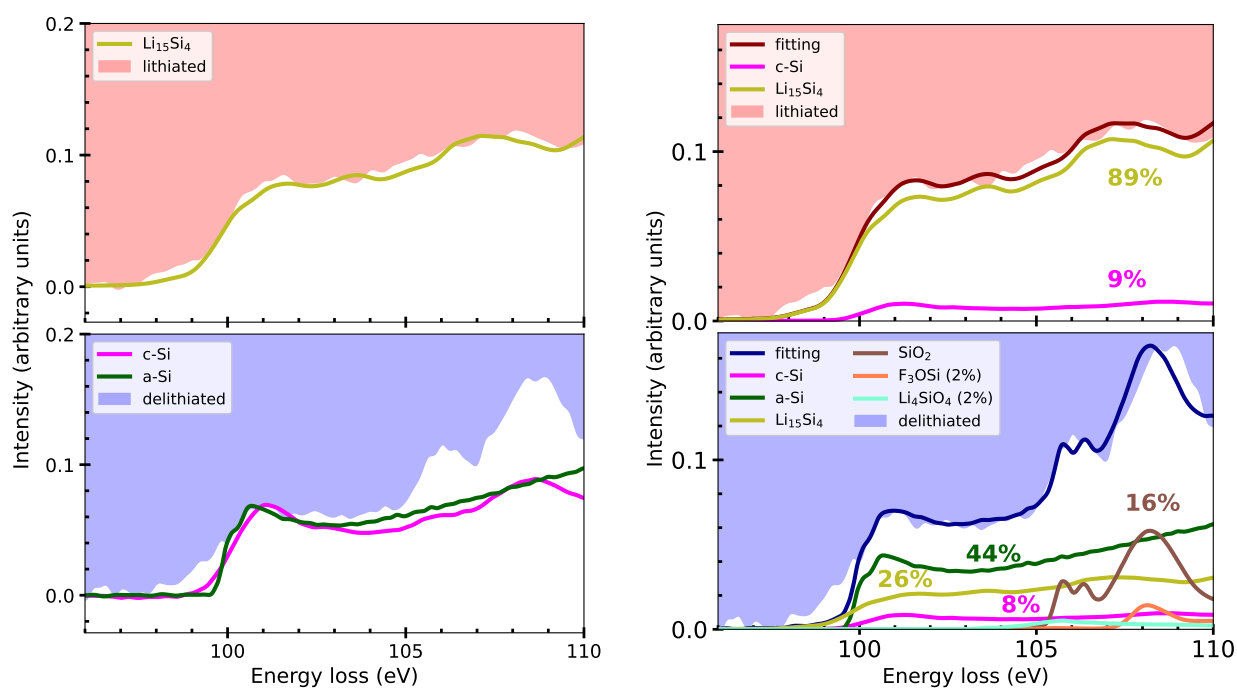

Figure S6: (Left) Comparison of the electrode Si $L_{2,3}$ edge spectrum in both SOC with the c-Si (this experiment), a-Si[3] (XANES experiment) and $Li_{15}Si_4$ (theory) references. (Right) LC fitting for the Si $L_{2,3}$ edge corresponding to the two SOC using as references c-Si, a-Si[3] and $SiO_2$[2] from experiments and $Li_{15}Si_4$, $F_3OSi$ and $Li_4SiO_4$ from theory.

## LC fitting for the Li K-edge

Figure S7 (left) shows the the comparison of the lithiated (top) and delithiated state (bottom) with the LC fitting model obtained by using the experimental references only (grey area),

7

that is, $Li_2CO_3$, LiF and $LiPF_6$. Specifically, here the least square fitting procedure fails, giving mostly $Li_2CO_3$ contribution (96% $Li_2CO_3$ only in the lithiated state and 86% $Li_2CO_3$ plus 8% LiF and 3% $LiPF_6$ in the delithiated one). By adjusting manually the LC model to try to reproduce the features between 60 and 64 eV and around 68 eV, we clearly see that the missing contributions are for the peak around 59 eV for both SoC (and the dip around 64 eV for the lithiated case).

The inclusion of the theory reference (right panels) for a prototypical Li-silicide ($Li_{15}Si_4$, green dashed line) gives a much better agreement of the LC model for the 100% SoC (dark red dashed line). In particular, the least square fit procedure is able to reproduce remarkably well the XRS data using 55% of $Li_2CO_3$ and 45% of $Li_{15}Si_4$. Adding manually few % of LiF adjusting the remaining % does not worsen the fitting. On the contrary, for the 0% SoC some contribution at low energy is still missing (dark blue dashed line).

## Estimate of the Li trapped in Si NPs

In the main text, we make use of recent titration-gas chromatography experiments[5] conducted on similar Si NPs with a Na-CMC binder and acetylene carbon black as the conductive additive to estimate the amount of Li trapped in Li–Si alloys. As a sanity check, we propose here an alternative approach based on galvanostatic measurements. We make use of the galvanostatic measurements presented in Ref. 4, where similar samples are held at decreasing low cutoff voltages (from 600 to 1 mV) and then fully delithiated at a cell potential of 2.0 V. The values of the recovered capacity for each lithiation potential, extracted from Figure 3a of Ref. 4, are reported in Figure S8 (purple triangles). By using a linear fitting of these values, we interpolate the capacity loss corresponding to the lithiation potential used in this work, i.e. 10 mV, which corresponds to 17% of the total capacity loss. This agrees quite well with the value obtained from titration-gas chromatography,[5] i.e. 19%.

8

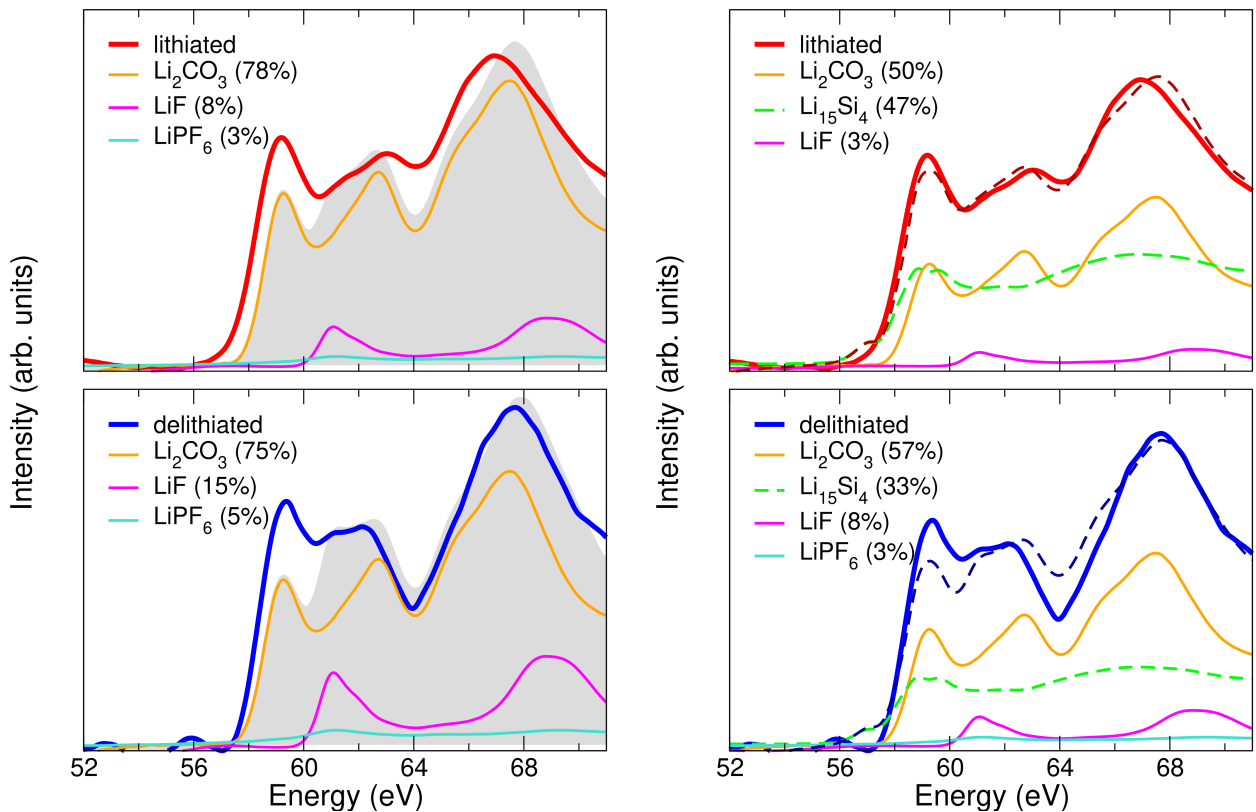

Figure S7: LC fitting for the Li K-edge corresponding to the two SOC (left, grey area) using the experimental references only ($Li_2CO_3$, LiF and $LiPF_6$) and (right, dark red/blue dashed lines) including both experimental and theory references ($Li_2CO_3$, LiF, $LiPF_6$ and $Li_{15}Si_4$, repsectively).

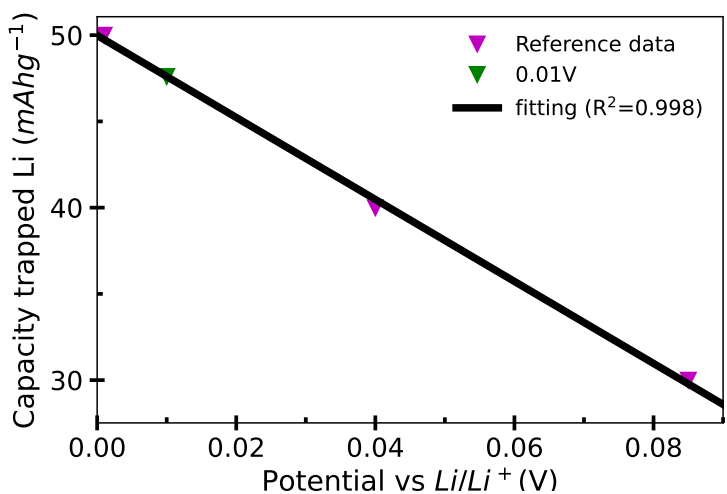

Figure S8: Linear fitting of the recovered capacity from Li trapped in Si NP (data taken from Ref. 4) vs. the final lithiation potential. The lithiation potential used in this work, 10 mV, implies that 17% of the total capacity loss corresponds to the Li trapped in Li-Si alloys.

9



\end{document}